\documentclass[journal=jctcce, manuscript=article]{achemso}
\usepackage[english]{babel}
\usepackage{siunitx}
\usepackage{graphicx}
\usepackage{amsmath}
\usepackage{bm}
\usepackage{braket}
\usepackage[utf8]{inputenc}
\usepackage[T1]{fontenc}
\usepackage{subcaption}
\usepackage{booktabs}
\usepackage{multirow}
\usepackage{amssymb}

\usepackage{todonotes}
\usepackage{color}

\usepackage{hyperref}

\newcommand{\artoo}{$R^2$ }

\title{Charge Transfer Simulations using Hamiltonian Elements and Forces from Neural Networks}

\author{Philipp M. Dohmen}
\affiliation{Institute of Physical Chemistry (IPC), Karlsruhe Institute of Technology, 76131 Karlsruhe, Germany}
\alsoaffiliation{Institute of Biological Interfaces (IBG2), Karlsruhe Institute of Technology, 76131 Karlsruhe, Germany}
\altaffiliation{These authors contributed equally to this work.}

\author{Mila Krämer}
\affiliation{Institute of Physical Chemistry (IPC), Karlsruhe Institute of Technology, 76131 Karlsruhe, Germany}
\altaffiliation{These authors contributed equally to this work.}

\author{Patrick Reiser}
\affiliation{Institute of Nanotechnology (INT), Karlsruhe Institute of Technology, 76131 Karlsruhe, Germany}

\author{Pascal Friederich}
\affiliation{Institute of Nanotechnology (INT), Karlsruhe Institute of Technology, 76131 Karlsruhe, Germany}
\alsoaffiliation{Institute of Theoretical Informatics (ITI), Karlsruhe Institute of Technology, 76131 Karlsruhe, Germany}

\author{Marcus Elstner}
\affiliation{Institute of Physical Chemistry (IPC), Karlsruhe Institute of Technology, 76131 Karlsruhe, Germany}
\alsoaffiliation{Institute of Biological Interfaces (IBG2), Karlsruhe Institute of Technology, 76131 Karlsruhe, Germany}

\author{Weiwei Xie}
\affiliation{Key Laboratory of Advanced Energy Materials Chemistry (Ministry of Education),
Renewable Energy Conversion and Storage Center, College of Chemistry, Nankai University, Tianjin, 300071, China}
\email{xieweiwei@nankai.edu.cn}

\begin{document}

\begin{abstract}
The trajectory surface hopping method has been widely used in the simulation
of charge transport in organic semiconductors. In the present study, we employ
the machine learning (ML) based Hamiltonian to simulate the charge transport
in anthracene and pentacene. The neural network (NN) based models are able to 
predict not just site energies and couplings but also the gradients of the site 
energy as well as off-diagonal gradients necessary for forces. We train the models 
on DFTB-quality data for both anthracene and pentacene. By using the obtained models 
in propagation simulations, we evaluate their performance in reproducing hole mobilities 
in these materials in terms of both quality and computational cost. The results show that 
the charge mobilities obtained using the NN-based Hamiltonian are in very good agreements with
the charge mobilities computed using the DFTB-based Hamiltonian.

\end{abstract}

\section{Introduction}
\label{sec:intro}

	Molecular organic semiconductors (OSCs) have become ubiquitous components of electronic devices, in the forms of organic light emitting diodes (LEDs) in modern display technologies \cite{armstrong2009_Organic, meerheim2009_Efficiency, kulkarni2004_Electron}, organic field effect transistors \cite{dimitrakopoulos2002_Organic, newman2004_Introduction, facchetti2007_Semiconductors}, or organic photovoltaic devices \cite{kippelen2009_Organic, dennler2009_Polymerfullerene, deibel2010_Polymer}.
	Compared to the silicon-based technologies for LEDs, transistors, or photovoltaic devices, OSCs are inexpensive and easier to manufacture and process \cite{dey2015_Organic}.
	With the near-limitless variety of organochemical compounds, searching for novel candidates with specific properties or optimizing known materials is a costly process.
	Here, theoretical approaches using simulation techniques can supplement or shorten experimental studies by efficiently screening large portions of chemical space for compounds with promising properties before committing to synthesizing them.
	This requires fast and robust methods for simulating charge and exciton transfer in large molecular systems which can reproduce experimentally observable quantities such as charge carrier mobilities.
	
	Simple models used to describe the transfer of charges and excitons assume either a band-like or hopping-like regime \cite{oberhofer2017_Charge}.
	In the band model, the charge is strongly delocalized as the couplings between the charge carriers are large and the reorganization energy is small.
	In contrast, hopping-like transfer occurs when the reorganization energy is large and the coupling is small, creating an activation barrier which is only overcome via thermal fluctuations.
	This results in a strong localization of the charge on single molecules, and transfer occurs in discrete hops.
	
	In OSCs, however, neither of these approximations hold \cite{oberhofer2017_Charge}, and direct simulations using quantum-chemical (QM) methods are needed to describe the transfer process properly.
	These approaches use non-adiabatic molecular dynamics (NAMD) methods and explicitly take into account the electronic degrees of freedom by solving the time-dependent Schrödinger equation (TDSE) \cite{kubar2013_Modeling, heck2015_Multiscale, kranz2016_Simulation, giannini2019_Quantum}.
	To mitigate the computational costs inherent in a quantum-chemical description of large systems along extensive molecular dynamics (MD) trajectories, the nuclei are treated classically and the electronic structure of the individual molecules is simplified.
	Using a conceptual coarse-graining of the transfer process, the electronic structure of the entire system can be compacted to two core parameters: the energies of frontier orbitals on each molecule and the pairwise coupling terms which describe their interaction.
	The response of a molecule to a change in its occupation can be taken into account explicitly by approximating the forces in the charged state.
	Alternatively, this relaxation can be modeled \emph{implicitly} by artificially reducing its energy by an empirically determined reorganization energy.
	
	Still, the costs of the \textit{ab initio} or semi-empirical models used to describe the electronic structure can be prohibitively large when the system size and trajectory lengths needed to accurately obtain observables are taken into account.
	Data-driven approaches, more specifically machine learning (ML) methods, can make use of the fact that the quantum-chemical calculations are highly repetitive, i.e. the conformations sampled during such simulations are very similar.
	By training the ML model on a small but representative data set for which the relevant electronic parameters have been calculated using a quantum chemistry method, the computationally simpler ML method could then be used to drive the transfer simulations.
	Several works have previously investigated the applicability of ML models to charge and exciton transfer properties:
	Both kernel ridge regression (KRR) \cite{lederer2019_Machine, wang2019_Machine, kramer2020_Charge} and neural network (NN) models \cite{musil2018_Machine, caylak2019_Evolutionary} have been used for predicting charge transfer couplings in organic and metal-organic systems.
	For excitonic properties, neural networks have been used to predict excitation energies in biological systems \cite{hase2016_Machine}, and there has been extensive work on ML-driven excited-state molecular dynamics \cite{li2021_Automatic, westermayr2019_Machine, westermayr2020_Machinea, westermayr2020_Combining} for small organic molecules.
	
	In our previous work \cite{kramer2020_Charge}, we used KRR models for predicting site energies and couplings in anthracene for both charge and exciton transfer applications.
	By using the fast and accurate density-functional tight binding (DFTB) method\cite{porezag1995_Construction, seifert1996_Calculations, elstner1998_Selfconsistentcharge, niehaus2001_Tightbinding, kranz2017_Timedependent, aradi2007_DFTB, hourahine2020_DFTB} to generate reference data cheaply and automating model training we created a scheme which was easily applicable to new systems without requiring individual models to generalize across chemical space.
	We then applied these models in propagation simulations, comparing the resulting observables with experimental or higher-level computational data.
	Our models were able to reproduce hole mobilities close to the reference and experimental values when using the mean-field Ehrenfest (MFE) propagator using the implicit relaxation scheme with a fixed reorganization energy.
	When using the Boltzmann-corrected fewest-switches surface hopping (BC-FSSH) method for propagation, we observed that while the ML method was close to the appropriate DFTB reference, both were far off from the values obtained if molecular relaxation was explicitly taken into account.
	
	While the implicit relaxation scheme has since been improved to give better mobilities, it is still an approximation which only has limited applicability in systems where the assumption that the molecules relax instantly does not hold.
	Therefore, obtaining the occupation-dependent forces required for molecular relaxation from a machine-learning model is an obvious next step.
	These forces can be considered as a correction to the neutral-state forces provided by the force-field (FF) driving the nuclear dynamics in the simulation and obtained as derivatives of the site energies from the Hamiltonian used for propagation with respect to the atomic positions.
	Unfortunately, obtaining a prediction of a property and its derivatives w.r.t. the inputs from a KRR model is tricky.
	Efforts to extend the KRR formalism to this end \cite{christensen2019_Operators} were unsuitable for our application due to the high computational costs they incur.
	
	In this work, we therefore present neural network based models which are able to predict not just site energies and couplings but also the gradients of the site energy necessary for occupation-dependent forces.
	We restrict ourselves to simulations of charge transfer and train the models on DFTB-quality data for both anthracene and pentacene.
	By using the obtained models in propagation simulations, we can evaluate their performance in reproducing hole mobilities in these materials in terms of both quality and computational cost.
		
\section{Methods}
\label{sec:methods}
		
	\subsection{Background}
	\label{ssec:methods:bg}

		\subsubsection{Charge Transfer Simulations}
		\label{sssec:methods:bg:qc}
		
			In order to make charge transfer simulations feasible, the degrees of freedom (DOF) which receive full quantum-mechanical treatment must be reduced as much as possible.
			In our simulations we therefore treat the nuclear degrees of freedom classically using Newton's equations of motion and a molecular mechanics (MM) force field.
			The electronic degrees of freedom are calculated quantum-mechanically and the time-dependent Schrödinger equation is used to calculate electronic dynamics\cite{kubar2013_hybrid}.
			To further reduce computational costs, we make use of the weakness of the non-covalent interactions between individual molecules, as the orbital structure of two molecules in proximity to each other can be assumed to be essentially identical to that of the isolated molecules.
			This approximation allows us to split the system into molecular fragments and obtain the electronic structure of each fragment in a separate QM calculation.
			The wave function for transport of the hole or electron is then expressed as a linear combination of orthogonal molecular orbitals (MOs) $\ket{\phi_m}$ localized on individual fragment molecules $A$:
			\begin{equation}\label{eq:holeWF}
				\Psi = \sum_{A}\sum_{m \in A} a_m \ket{\phi_m}
			\end{equation}
			This is called the fragment molecular orbital (FMO) approach.
			Only the frontier MOs ($m$) of the fragments are considered\cite{heck2015_Multiscale}, as charge transfer in OSCs occurs in a narrow energy band around the Fermi level.
			The energies of these frontier MOs for each fragment and the pairwise couplings between them can be used to construct a coarse-grained Hamiltonian for propagation of the electronic DOF \cite{kitaura1999_Fragment, kubar2008_Efficient} as
			\begin{equation}
				H^0_{mn} = \bra{\phi_m}\hat{H}[\rho_0]\ket{\phi_n} \; ,
			\end{equation}
			where $\hat{H}[\rho_0]$ is obtained from the QM calculations of the fragments.
			The \emph{diagonal} elements of the coarse-grained Hamiltonian are on-site-terms (site energies), while its \emph{off-diagonal} elements are pairwise couplings between the fragments.
			They can be obtained with only a slight increase in computational cost, involving only the relevant frontier orbitals.
			The Hamiltonian can then be used in one of the commonly used propagation algorithms, such as the mean-field Ehrenfest method \cite{ehrenfest1927_Bemerkung, mclachlan1964_variational} (MFE) or variations of the fewest-switches surface hopping \cite{tully1990_Molecular, carof2017_Detailed} (FSSH) method to solve the coupled equations of motion for electronic and nuclear DOF.
			
			In this work, we used a variant of FSSH using a Boltzmann correction instead of non-adiabatic coupling vectors for velocity re-scaling.
			In the FSSH method, the electronic wave function is expressed as a linear combination of adiabatic basis functions $\{|\psi_i \rangle\}$:
			\begin{equation}\label{adiabexpan}
				\Psi = \sum C_i^\mathrm{ad} |\psi_i \rangle \;,
			\end{equation}
			where $C^{ad}_{i}$ are the expansion coefficients of the electronic wave function
			in the adiabatic representation.
			Inserting Eq.~\ref{adiabexpan} into the time-dependent Schrödinger equation and projecting onto the adiabatic electronic basis states, we obtain
			\begin{equation}\label{eq:tdse2}
				i \hbar \frac{d C_j^\mathrm{ad}}{d t} = C_j^\mathrm{ad} H^\mathrm{ad}_{j} - i \hbar \sum_{k}
				C_k^\mathrm{ad} \boldsymbol{\dot{R}} \boldsymbol{d}_{jk}  \;,
			\end{equation}
			where $\boldsymbol{d}_{jk}$ denotes the non-adiabatic coupling vector (NCV) and $H^\mathrm{ad}_{j} $ is the potential energy (PE) of adiabatic state $j$.
			The diagonalization of the matrix $\boldsymbol{H^0}$ yields the adiabatic potential energies
			\begin{equation}\label{diag}
				\boldsymbol{H^{ad}} = \boldsymbol{U^{\dagger}}\boldsymbol{H^0}\boldsymbol{U}\;,
			\end{equation}
			where $\boldsymbol{U}$ is the diabatic-to-adiabatic transformation matrix.
			
			Classical trajectories are propagated on a single electronic state according to Newton's equations of motion:
			\begin{eqnarray}\label{cleom1}
				m_k \ddot{R}_k  &=& - \frac{\partial E_\mathrm{MM}^\mathrm{tot}}{\partial R_k} + \frac{\partial H^\mathrm{ad}_j}{\partial R_k} - \frac{\partial \Delta E^\mathrm{QM/MM}}{\partial R_k} \nonumber \\
								&=&  - \frac{\partial  E_\mathrm{MM}^\mathrm{tot}}{\partial R_k} + \sum_{mn} U_{am} U_{an} \frac{\partial H^0_{mn}}{\partial R_k} - \frac{\partial \Delta E^\mathrm{QM/MM}}{\partial R_k} \;,
			\end{eqnarray}
			where $H^\mathrm{ad}_j$ denotes the adiabatic energy of the current state $j$.
			
			The hopping probability from the current state $j$ to another state $k$ is defined as
			\begin{eqnarray}\label{eq:hopprob}
				P_\mathrm{FSSH}^{j \rightarrow k} &=& \max \left\{ 0, \frac{2 \boldsymbol{\dot{R}} \boldsymbol{d}_{jk}\mathrm{Re} \left(C_{k}^\mathrm{ad*} C_{j}^\mathrm{ad}\right)}{|C_{j}^\mathrm{ad} |^{2}} \Delta t \right\}  \;.
			\end{eqnarray}
			A uniform random number $\xi$ between $0$ and $1$ is generated and a switch from state $j$ to state $k$ takes place, if
			\begin{equation}\label{eq:hopcondi}
				\sum_{i=1}^{k-1} P_\mathrm{FSSH}^{j \rightarrow i} < \xi \leq \sum_{i=1}^{k} P_\mathrm{FSSH}^{j \rightarrow i} \;.
			\end{equation}
			
			To ensure the conservation of total energy for each trajectory, the nuclear momenta must be adjusted when a hop occurs.
			This is usually achieved by re-scaling the nuclear velocities to compensate the change of the PE.
			While distributing this effect isotropically across all velocity components is possible, better results are obtained if the velocities are re-scaled in the direction of the non-adiabatic coupling vectors.\cite{carof2017_Detailed}
			If the adjustment of momentum is not sufficient to compensate the change in PE (an `energy-forbidden' or `frustrated' hop), the trajectory continues in the original electronic state $j$, and the momenta along the NCV are reversed.
			
			An alternative to using NCVs is to re-scale the hopping probability using the Boltzmann factor $\exp\left(-\frac{H^{\mathrm{ad}}_k- H^{\mathrm{ad}}_j}{k_B T}\right)$\cite{akimov2013_PYXAID} and leave the velocities unchanged.
			This Boltzmann-corrected FSSH method has the advantage that computationally demanding NCVs are not necessary, while giving better results than isotropic re-scaling.
			
			After a hop occurs, fragments occupied by the charge carrier must respond to the changed electronic state by relaxing their geometry, which reduces their site energies.
			As mentioned above, this can either be accomplished by artificially lowering the site energy by an empirically determined value for the reorganization energy, or by applying the forces resulting from the charged state.
			As the force field which governs nuclear dynamics is parameterized to reproduce the forces in the neutral state, the forces for occupied fragments can be adjusted using the difference between forces in the neutral and charged states.
			Note that this approach is different from the adiabatic forces used in other non-adiabatic dynamics methods.
			
			Under the approximation that the electronic structure of the charged fragments is identical to that of the neutral molecules except for the occupations of the individual orbitals we can obtain the difference between charged and neutral state forces as the forces resulting from only the frontier orbitals in which the occupations change.
			These \emph{FMO}-forces are the exact derivatives of the diagonal and off-diagonal terms in the CT Hamiltonian.
			As the coupling with neighboring fragments only results in small force contributions compared to the FF and diagonal terms, we neglect the off-diagonal derivatives in this work.
			In the future, the off-diagonal derivatives can be used for both relaxation and calculating non-adiabatic coupling vectors.			
			
			Another way to obtain the force correction necessary for relaxation is to calculate the forces from the charged molecule's electronic state in addition to the neutral-state calculation and form the difference between the forces in the two states.
			\begin{equation}
			F^{\Delta} = \frac{\partial}{\partial R_k}\,E^{\mathrm{charged}} - \frac{\partial}{\partial R_k}\,E^{\mathrm{neutral}}
			\,.
			\end{equation}
			This \emph{$\Delta$}-force approach however requires an additional QM calculation of the charged fragment, which is why it is not usually used during propagation simulations.
			
			In all systems examined in this work, the occupation-dependent QM forces were approximately one order of magnitude smaller than the FF baseline.
			This is quite helpful when replacing the QM method with an ML model, as it provides a larger margin of error for predictions before they significantly impact the stability of the simulation by creating physically unreasonable geometries.

		\subsubsection{Machine Learning Methods}
		\label{sssec:methods:bg:ml}

			Neural networks are non-linear statistical models which are widely used to solve complex regression or classification problems.
			A feed-forward neural networks takes a vector of inputs $(x_1, x_2, \dots x_i)$ and makes a number of linear combinations $(n_1, n_2, \dots n_j)$ of these inputs referred to as neurons.
			The weights $w_{ij}$ determine the contribution of each input $i$ to each combination $j$.
			The result of each linear combination $n$ is then passed through a non-linear activation function, which allows the neural network to approximate arbitrarily complex functions.
			These results are then combined to make another layer of linear combinations, and the process can be repeated until the desired network depth is reached.
			The final layer of the network then carries the results, i.e. the prediction the network makes for a given input.
			
			All weights in the NN are initialized randomly and adjusted such that the prediction for inputs for which the result value is known matches the reference value as closely as possible as measured by a metric referred to as \emph{cost} function.
			Cost functions can be as simple as the mean squared error (MSE) between prediction or reference, but can also include other terms to impose additional constraints on the structure of the network.
			Training is then the process of minimizing the cost function by giving the network inputs from the training data set and evaluating the cost.
			Minimization uses the backpropagation algorithm, in essence taking the derivative of the cost function w.r.t. all weights in the network (employing the chain rule for deep networks with multiple layers) repeatedly and making small steps towards the minimum of the function.
			
			The choices made in the construction of the network (e.g. how many layers and neurons to use) can strongly influence the performance of the resulting models.
			As these parameters are not optimized during the training itself, the proper values for them must be determined by another process, and they are referred to as \emph{hyperparameters}.
			Hyperparameter optimization usually requires repeated training of NNs on the same data set while the values for the hyperparameters are varied.
			This is usually computationally demanding, as the search space can grow rather large very quickly, and can be performed using various optimization algorithms.

	\subsection{Computational Details}
	\label{sec:methods:det}
		
		\subsubsection{Generation of training data}
		\label{sssec:methods:det:gen-data}
		
			The procedures for generating training data were identical for both the anthracene and pentacene systems.
			For each system, we created a crystal super-cell based on experimental crystal structures\cite{mason1964_crystallography}.
			The anthracene super-cell contained 8 $\times$ 8 $\times$ 5 molecules along the crystal axes (\textit{a}, \textit{b} and \textit{c}), respectively.
			In pentacene, the size of the crystal was 8 $\times$ 16 $\times$ 5 molecules.
			
			Force field parameters for each system were derived from the general AMBER force field (GAFF) \cite{wang2004_Development, wang2006_Automatic}.
			Atomic charges were generated from the restrained electrostatic potential (RESP) fitting procedure \cite{singh1984_approach, besler1990_Atomic}, calculated at HF/6-31G* \cite{petersson1988_complete, petersson1991_complete} level of theory using \texttt{Gaussian09} \cite{frisch2009_Gaussian}.
			
			After an initial energy minimization the temperature was equilibrated for \SI{1}{ns} with a time step of \SI{2}{fs} using the Nose-Hoover thermostat \cite{evans1985_Nose} at \SI{500}{K}.
			Subsequently, productive MD simulations were run for \SI{1}{ns} with a time step of \SI{1}{fs}, in which structures were saved every \SI{1}{ps}.
			All MD simulations were performed with the \texttt{GROMACS} 4.6 software package \cite{berendsen1995_GROMACS, abraham2015_GROMACS}.
			
			As the gradients obtained from the ML methods will be used to relax molecular geometries, it is crucial that gradient predictions are reliable across the entirety of the potential energy surfaces (PES) accessible throughout the simulation.
			In order to keep the learning procedure and implementation of ML predictions into the simulation code simple, we refrained from using active learning or similar techniques.
			Instead, we constructed the training data set by sampling geometries along both the PES of the neutral system and the charged system at a temperature higher than that used during the productive simulations for obtaining mobilities.
			One standard MD simulation was done to sample geometries on the neutral PES, while multiple separate NAMD simulations were performed to sample on the PES of positively charged molecules.
			Here, the charge was constrained to one molecule.
			The final data set then contained data from both neutral and charged PES in a 50:50 ratio.
			
			Eventually, single molecule structures and pair-structures sampled in the neutral and charged states were randomly selected from a 5 $\times$ 5 $\times$ 3 molecular cube in the center of the crystal for subsequent calculations of site energies, their derivatives and couplings.
			The HOMO was chosen as the frontier orbital for hole transport.
			Site energies and electronic couplings were calculated using non-self-consistent DFTB \cite{porezag1995_Construction, seifert1996_Calculations}, often referred to as DFTB1.
			For the derivatives of the site energy, different data sets were constructed corresponding to the two approaches to obtain the gradients discussed in \autoref{sssec:methods:bg:qc}.
			The \emph{FMO} gradients are obtained as gradients of the HOMO of the neutral molecule using DFTB1.
			In contrast, the \emph{$\Delta$} gradients were obtained as the difference between the total gradients of the neutral and charged molecules using self-consistent-charge DFTB \cite{elstner1998_Selfconsistentcharge} with a long-range corrected functional \cite{niehaus2012_Range} (LC-DFTB2) as implemented in the \texttt{dftb+} (version 19.1) \cite{aradi2007_DFTB, hourahine2020_DFTB} program.
			We used LC-DFTB2 for this data set, as the long-range corrected functional can correct the underestimation of the reorganization energies observed when using the GGA-derived non-self-consistent DFTB.
			Additionally, gradients for anthracene are obtained from DFT calculations with the B3LYP \cite{becke1993becke} and $\omega$B97X \cite{chai2008systematic} functionals, employing the def2-TZVP \cite{weigend2005balanced} basis-set together with the def2/J \cite{weigend2006accurate} auxiliary basis as implemented in \texttt{orca} (version 5.0.1) \cite{neese2012orca, neese2018software}.
			
		\subsubsection{Training of Machine Learning Models}
		\label{sssec:methods:det:training}
		
			We adapt the neural network architecture previously presented by \citeauthor{li2021_Automatic} \cite{li2021_Automatic} for our application, as it has been constructed in order to predict energies and forces electronic states of small organic molecules.
			The network uses as inputs the spatial coordinates of the atoms in the system, from which inverse interatomic distances between atoms are calculated as a translationally and rotationally invariant representation.
			In models predicting the site energies and their derivatives, all intramolecular interatomic distances are included in the representation.
			For the coupling models, we use only the intermolecular block of the matrix of inverse distances as this has been shown to be superior for predicting coupling elements\cite{wang2019_Machine} and reduces the computational cost of each prediction\footnote{Our initial experiments confirmed that the reduced representation gives lower errors at all training set sizes}.
			The trainable part of the network is a multi-layer dense feed-forward NN using the leaky softplus activation function ($f(x) = (1-\alpha) \cdot \mathrm{softplus}(x) + \alpha \cdot x$) \cite{zhao2018_novel} with a slope $\alpha = 0.03$.
			
			Models were trained using the \texttt{pyNNsMD} package \cite{li2021_Automatic} and Tensorflow 2.3 \cite{martinabadi2015_TensorFlow} with the Keras API \cite{chollet2015_Keras}.
			Weights were optimized using the Adam optimizer\cite{kingma2017_Adam} and mean-squared-error loss.
			All models received training and validation data in a 9:1 ratio, and the loss on the validation set was monitored every epoch, so training could be stopped when it did not improve for more than 100 epochs (early stopping).
			Feature calculation and scaling are both implemented as layers of the neural network, reducing implementation overhead when using the models for prediction.
			While training the models for the couplings was straightforward, during training of the site energy models the loss was calculated on both the network's prediction and the gradients w.r.t. input coordinates.
			The relative weights of these two parts of the loss function were considered a hyperparameter of the model.
			
			We used the Hyperband algorithm as an efficient and automatable way to find working combinations of hyperparameters\cite{li2018_Hyperband}.
			However, the quality of the resulting models was only weakly dependent on the chosen hyperparameters, indicating that a thorough hyperparameter search is unnecessary and can be omitted when applying the method to new systems.
			More details on the network configuration, training and hyperparameter search can be found in the supporting information.
			
		\subsubsection{Charge Transfer Simulations}
		\label{sssec:methods:det:simulation}

		We generated separate crystals for the transfer simulations containing \num{40} $\times$ \num{30} $\times$ \num{14} (\num{36}) molecules for anthracene in \textit{a}-direction, \num{20} $\times$ \num{40} $\times$ \num{5} (\num{36}) molecules for anthracene in \textit{b}-direction and \num{42} $\times$ \num{84} $\times$ \num{3} (\num{73}) molecules for pentacene in \textit{T1}-direction.
		We chose one-dimensional lines of molecules along the respective directions in the middle of each crystal for the QM zone, with the number of fragments specified in brackets.
		After equilibration at \SI{300}{K} structures in equidistant time intervals of \SI{10}{ps} were chosen as starting structures for subsequent simulations of charge transfer.
		The hole wave function was initially localized on the first molecule.
		We used a time step of \SI{0.1}{fs} for the propagation of nuclei, the TDSE was integrated numerically with the fourth-order Runge-Kutta algorithm with an integration time step of \SI{0.01}{fs}.
		Averages of observables were calculated on \num{1000} trajectories, which were simulated for \SI{1}{ps} each, while the first \SI{0.35}{fs} were regarded as initial equilibration of the charge.
		All charge transfer simulations were performed within a local version of GROMACS 4.6 were DFTB, NN as well as the BC-FSSH methods were implemented.

		We calculated the hole mobility $\mu$ with the Einstein-Smoluchowski equation \cite{kubo2012statistical} as
		\begin{equation}\label{eq:mobility-einstein}
		\mu = \frac{e D}{k_\mathrm{B} T}\;,
		\end{equation}
		where $e$ is the elementary charge, $k_\mathrm{B}$ is the Boltzmann constant and $T$ denotes the temperature.
		The diffusion coefficient $D$ \cite{frenkel2002understanding} is
		\begin{equation}\label{eq:diffcons-charge}
		D = \frac{1}{2n} \lim_{t \rightarrow \infty} \frac{d \, \mathrm{MSD}			(t)}{d t}\;,
		\end{equation}
		where $n$ is the dimensionality ($n=1$ for a one-dimensional chain).
		The mean square displacement (MSD)  of the charge across 			$N_{\text{traj}}$ trajectories is then
		\begin{equation}\label{eq:msd}
		\mathrm{MSD} (t) = \frac{1}{N_\mathrm{traj}}\sum_{l}^{N_\mathrm{traj}} 		\sum_{A}  (x_A(t)^{(l)} - x_0^{(l)})^2 P_A^{(l)} (t) \;,
		\end{equation}
		where $x_A(t)^{(l)}$ and $P_A (t)^{(l)}$ are the center of mass of molecule $A$ and the corresponding diabatic population along the trajectory $l$, respectively.
		$x_0^{(l)}$ is the center of charge at $t = 0$.

\section{Results}
\label{sec:res}

	\subsection{Model Training and Evaluation}
	\label{ssec:res:static}
	
		After training the models as described in \autoref{sssec:methods:det:training}, we evaluated how well they were able to predict held-out test data from their respective reference data set.
		All trained models converged well within two thousand epochs, with prediction errors shrinking as training set sizes increased.
		In contrast to the KRR models used in our previous work, the NN models were able to learn the sign of the couplings even at low training set sizes.
		Overall, the hyperparameters included in the search had only a negligible influence on the quality of models obtained compared to the statistical noise from the initialization of the weights.
		While it is possible that specific hyperparameter configurations might also reduce the number of epochs needed until convergence, the added computational cost of the hyperparameter search far exceeds these gains.
		We therefore conclude that for this application, a full hyperparameter search can be omitted.
		
		Here, we summarize the evaluation results for the models which we used in the propagation simulations in \autoref{ssec:res:dyn}.
		These models were obtained using a training set size of \num{30000} structures each, with \num{3333} additional structures used for validation and the optimal hyperparameter configurations resulting from the hyperparameter search.
		Metrics were evaluated on \num{10000} structures not used during training.
		The full results of the hyperparameter search and evaluation of individual models can be found in the supporting information.
			
		\subsubsection{Anthracene}
		\label{sssec:res:static:ant}
					
\begin{figure}
				\centering
				\includegraphics[width=\linewidth]{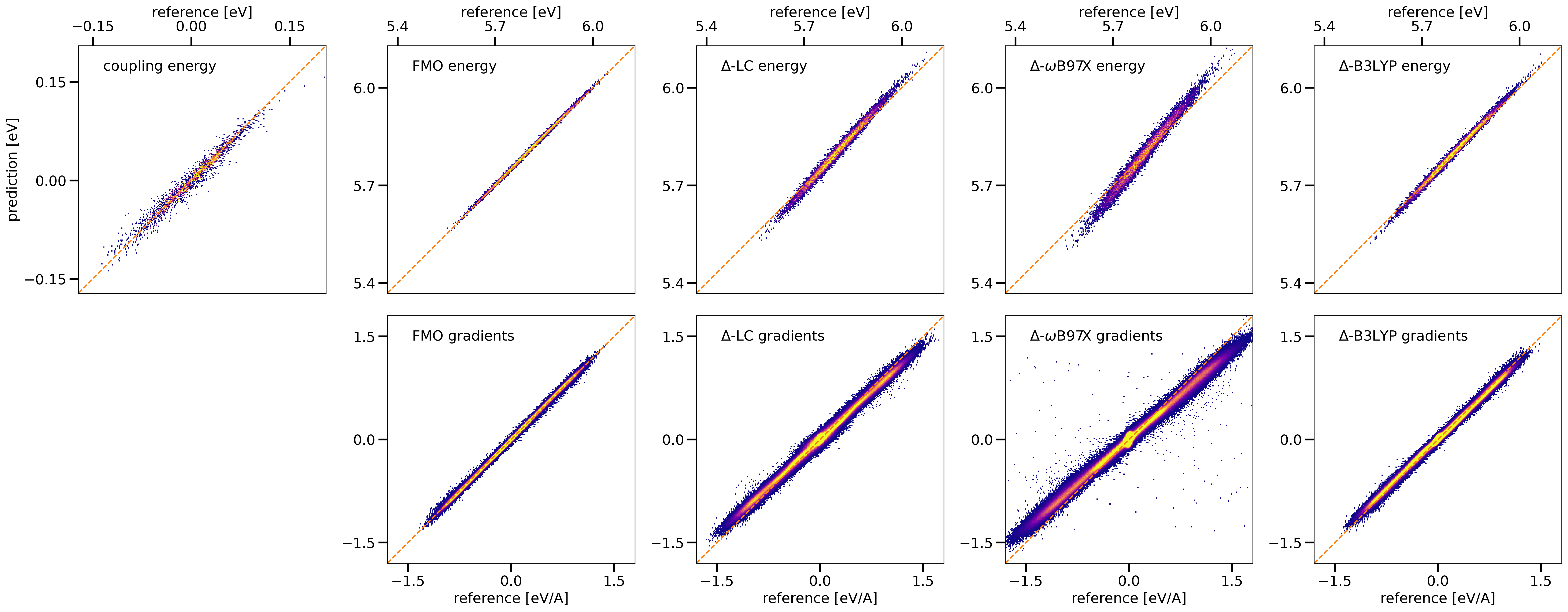}
				\caption{Two-dimensional histogram of model predictions vs. DFTB target for \num{10000} structures in Anthracene. Bright colors indicate high data density, unoccupied areas shown white.}
				\label{fig:res:ant-scatter}
			\end{figure}

			\autoref{tab:metrics-ant} gives a few crucial metrics for the models trained on the anthracene data.
			All models give excellent predictions with few outliers, as can be seen in \autoref{fig:res:ant-scatter}.
			\begin{table}
				\sisetup{round-mode=places, round-precision=3}
				\begin{tabular}{r  c c c c c}
				\toprule
		                                         &                         &              \multicolumn{2}{c}{FMO}              &          \multicolumn{2}{c}{$\Delta$-LC}            \\
		   \cmidrule(lr){3-4} \cmidrule(lr){5-6} & coupling                & site energy             & gradient                & site energy              & gradient                 \\\midrule
MAE $\left[\si{meV}(/\si{\angstrom})\right]$     & \num{1.77156331483275}  & \num{2.55350768566132}  & \num{10.580594651401}   & \num{9.467258}           & \num{41.130308}          \\
max err $\left[\si{meV}(/\si{\angstrom})\right]$ & \num{48.3487248420715}  & \num{18.2070732116699}  & \num{257.771849632263}  & \num{50.091267}          & \num{533.8345}           \\
					                       $R^2$ & \num{0.955376958001202} & \num{0.998022221423896} & \num{0.998467193738213} & \num{0.9747811405884006} & \num{0.9863981856166795} \\
				\bottomrule
				\end{tabular}
				\caption{Quality metrics for anthracene models.}
				\label{tab:metrics-ant}
			\end{table}
			The model for the couplings in anthracene performed quite well, reaching low mean absolute errors (\SI{1.77156331483275}{meV}) with an \artoo score of \num{0.955376958001202}.
			For the FMO site energies and their gradients, the model reached a MAE of \SI{2.55350768566132}{meV} for the energies and \SI{10.580594651401}{meV\per\angstrom} for the gradients, with \artoo scores above \num{0.998} for both.
			The model trained on the $\Delta$-LC data gave slightly worse results, and in contrast to the model for the FMO data did not perform equally well for both energies and gradients.
			This is a direct result from the construction of the LC data set, where LC-DFTB is only used to calculate the gradients, and these are learned in conjunction with the FMO site energies.
			
			The errors for site energies and couplings are comparable to those obtained for similar training set sizes in our previous work, indicating that the models should be sufficiently accurate to give good mobilities in simulations.
			The effects of the error on the forces cannot be so easily quantified, but the maximum prediction errors can be an indication whether the predicted forces could impede the stability of the simulation.
			The maximum prediction errors for both the FMO (\SI{0.257771849632263}{eV/\angstrom}) and $\Delta$-LC gradients (\SI{0.5338345}{eV/\angstrom}) are well below the forces needed to break covalent bonds ($\approx$\SIrange{1}{2}{eV\per\angstrom}\cite{grandbois1999_How}).
			Overall, these large errors are only seen for a few individual outliers, as less than one percent of gradient predictions show errors above \SI{40}{meV/\angstrom} (FMO) or \SI{160}{meV/\angstrom} ($\Delta$-LC).
			These metrics indicate that the gradient predictions should be sufficiently reliable for performing molecular dynamics.
		
		\subsubsection{Pentacene}
		\label{sssec:res:static:pen}
		
\begin{figure}
				\centering
				\includegraphics[width=\linewidth]{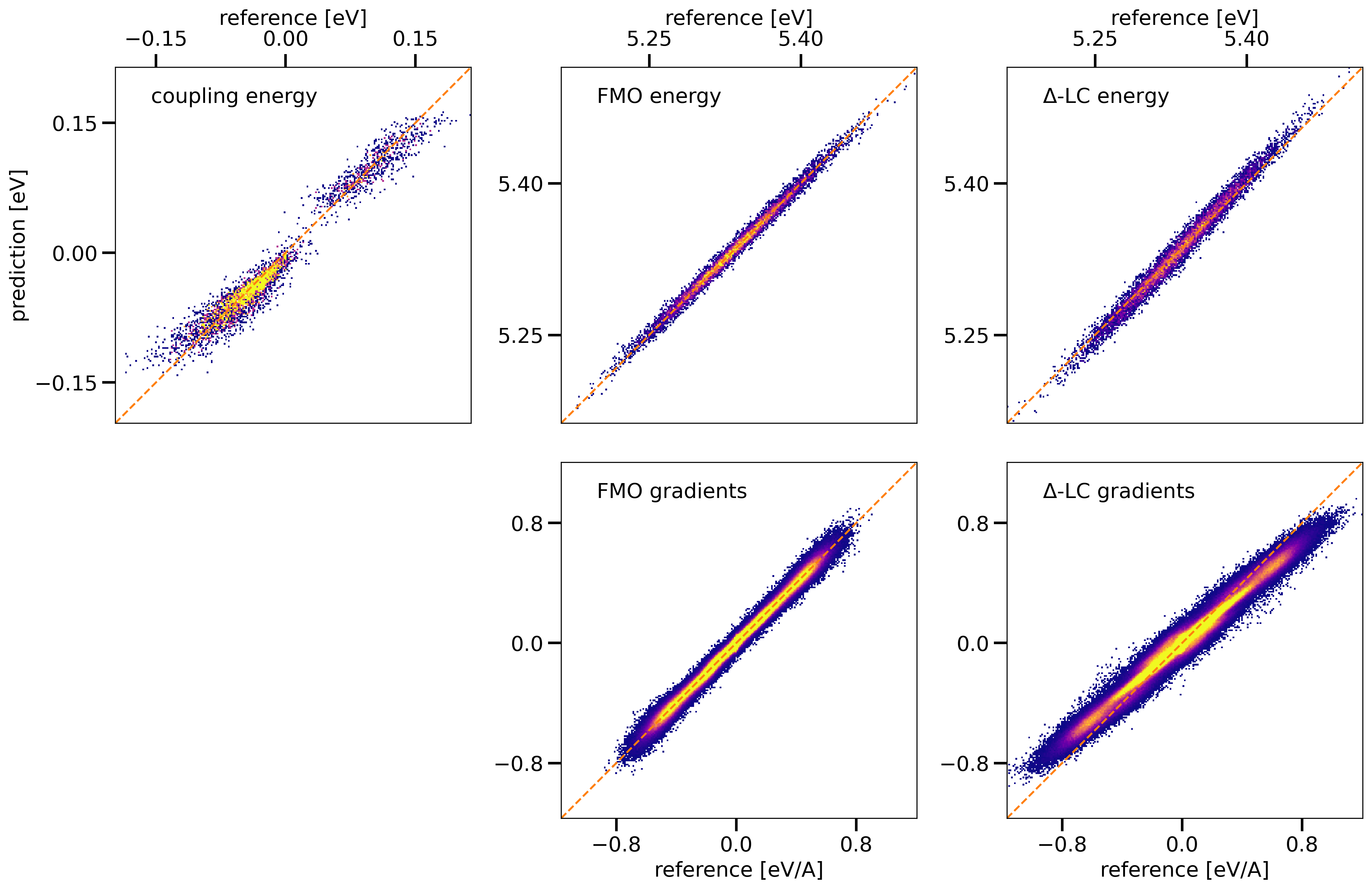}
				\caption{Two-dimensional histogram of model predictions vs. DFTB target for \num{10000} structures in Pentacene. Bright colors indicate high data density, unoccupied areas shown white.}
				\label{fig:res:pen-scatter}
			\end{figure}

			In pentacene, the results were quite similar and are summarized in \autoref{tab:metrics-pen} and visualized in \autoref{fig:res:pen-scatter}.
			In Pentacene, the couplings along the different crystal directions are quite dissimilar, leading to the bimodal distribution seen in the upper left of \autoref{fig:res:pen-scatter} and a more difficult learning target for the ML model.
			\begin{table}
				\sisetup{round-mode=places, round-precision=3}
				\begin{tabular}{r  c c c c c}
					\toprule
					&                 &             \multicolumn{2}{c}{FMO}              &          \multicolumn{2}{c}{$\Delta$-LC}          \\ \cmidrule(lr){3-4} \cmidrule(lr){5-6}
					&   coupling                      & site energy             & gradient                & site energy             & gradient                \\					\midrule
					MAE $\left[\si{meV}(/\si{\angstrom})\right]$ & \num{5.52033307030797}  & \num{3.43426736071706}  & \num{17.0863326638937}  & \num{5.54915703833103}  & \num{41.0026907920837}  \\
					max err $\left[\si{meV}(/\si{\angstrom})\right]$ & \num{68.0754259228706} & \num{22.456169128418} & \num{239.180862903595} & \num{31.646728515625} & \num{414.126932621002} \\
					$R^2$ & \num{0.95833341843381} & \num{0.98995666091023} & \num{0.988385120480334} & \num{0.974452069809537} & \num{0.958650955554732}\\
					\bottomrule
				\end{tabular}
				\caption{Quality metrics for pentacene models.}
				\label{tab:metrics-pen}
			\end{table}		
			As the mean magnitude of couplings is in the pentacene data set is \SI{28.2082818448544}{meV} compared to \SI{7.84547720104456}{meV} in anthracene, the MAE of the coupling predictions in pentacene is greater.
			The highly similar \artoo scores between the two systems indicate that the models have captured the data set equally well.
			The prediction errors for site energies and their gradients in the pentacene system were slightly worse than for the anthracene models, but still excellent.
			Maximum prediction errors for the gradients were slightly lower than in anthracene, and the error distribution remained narrow.

	\subsection{Machine-Learning Driven Simulations}
	\label{ssec:res:dyn}

		\subsubsection{Accuracy of Obtained Observables}
		\label{sssec:res:dyn:mobilities}

			In the following, we use the trained neural network models to perform NAMD simulations of hole transfer along linear chains of anthracene molecules in \textit{a}- and \textit{b}-direction as well as pentacene molecules in \textit{T1}-direction.
			These simulations are analyzed in terms of accuracy in reproducing hole mobilities compared to the reference method (DFTB) and experimentally determined values.
			
			In order to perform stable NAMD simulations, where the nuclear dynamics are partly driven by machine learned models, it turned out to be necessary to sample geometries for training data on the PES of both neutral and charged molecules.
			First tests with models trained on geometries in the neutral state only led to unstable/crashing simulations.
			Adding geometries that were sampled in the charged state gave robust models and stable simulations, such that no catastrophically bad predictions occurred in the more than \num{100} million simulation time steps performed for this work.
			As the occupation-dependent force corrections are approximately one order of magnitude smaller than the forces from the MM force field, the margin of error that the ML models can make before simulation stability is impacted is quite large and no further efforts (e.g. active learning) were necessary.

			The NAMD simulations presented here are driven by models, which are trained on the same diagonal energies but different gradients for relaxation, calculated either with FMO-DFTB1 (FMO), $\Delta$-LC-DFTB2 ($\Delta$-LC), $\Delta$-DFT/B3LYP ($\Delta$-B3LYP) or $\Delta$-DFT/$\omega$B97X ($\Delta$-$\omega$B97X).
			All ML driven simulations use the same molecule specific models for the prediction of off-diagonal energies.
			Additionally, DFTB driven simulations are performed as our reference method, using FMO-gradients for the relaxation.
			All other approaches are not implemented and computationally too demanding.
			The calculated hole mobilities for all investigated systems are displayed in Table \ref{tab:mobi-DFTBvsML}, which are computed from the corresponding averaged time-dependent MSD

			\begin{table}[ht]
				\sisetup{round-mode=places, round-precision=1}
				\caption{Calculated and experimental hole mobilities (in $\frac{cm^2}{Vs}$) for anthracene (\textit{a}- and \textit{b}-direction) and pentacene (\textit{T1}-direction).
				Simulations are driven by Hamiltonians and diagonal gradients from DFTB or NN models, the latter are trained on different diagonal gradients from DFTB (FMO, $\Delta$-LC) and DFT ($\Delta$-B3LYP, $\Delta$-$\omega$B97X), respectively.}
				\label{tab:mobi-DFTBvsML}
				\begin{center}
					\begin{tabular}{c c r r r}
						\toprule
						Method & Relaxation & ANT-a & ANT-b & PEN-T1 \\
						\midrule
						DFTB & FMO                     & \num{3.2}  & \num{8.4}  & \num{11.6}  \\
						\midrule			
						\multirow{4}{*}{NN} & FMO      & \num{2.9}  & \num{8.7}  & \num{14.6}  \\
						& $\Delta$-LC                  & \num{2.7}  & \num{8.1}  & \num{14.2}  \\
						& $\Delta$-B3LYP               & \num{2.5}  & \num{8.1}  & -           \\
						& $\Delta$-$\omega$B97X        & \num{1.9}  & \num{5.8}  & -           \\
						\midrule
						\multicolumn{2}{c}{Experiment} & \num{1.1} \cite{karl2001_experiment}  & \num{2.9} \cite{karl2001_experiment} & \num{10.5} \cite{jurchescu2004effect}  \\
						\bottomrule
					\end{tabular}
				\end{center}
			\end{table}

			Comparing mobilities from NN model driven simulations with the those from DFTB reference simulations (both using FMO forces), we find a good agreement for anthracene with deviations of \SI{9}{\%} and \SI{4}{\%} for \textit{a}- and \textit{b}-direction, respectively.
			The results for pentacene differ more, showing an overestimation of the mobility by roughly \SI{26}{\%} by the NN driven simulations.
			However, both the reference and the ML values are too large relative to experimentally determined results.
			Employing NN models trained on gradients from higher level DFTB or DFT, that would lead to higher reorganization energies, shows the expected trend of decreasing the mobilities.
			$\Delta$-LC and $\Delta$-B3LYP lead to almost the same reorganization energies, which is also reflected in the mobility values.
			For the small (ANT-a) and medium (ANT-b) mobility cases, results with $\Delta$-$\omega$B97X gradients are improved but the mobility is still overestimated by a factor of approximately \num{2}.
			In the high mobility case of pentacene the mobility is already much closer to the experiment using $\Delta$-LC forces, which lead to an overestimation of \SI{30}{\%}.

		\subsubsection{Comparison of Computational Costs}
		\label{sssec:res:dyn:costs}		
		
			Here we assess the performance of the learned models in respect of simulation time.
			To compare timings with the different methods and models we ran \num{100} trajectories with only two molecules for \num{1000} steps of NAMD simulation with the exact same settings compared to our previous calculations.
			The timings are estimated as averages of time spent on single calculations of site energies plus respective gradients and for couplings.
			For gradients in the $\Delta$ approach timings were averaged on the training-data calculations.
			All computations are performed on single CPU cores of Intel Xeon Silver 4214 @ \SI{2.20}{GHz} processors.

			\begin{table}[ht]
				\caption{Comparison of timings (in ms) for the calculation of diagonal Hamiltonian elements (site energies) plus gradients in anthracene and pentacene with DFTB, DFT and NN models.}
				\label{tab:timings-diag}
				\begin{center}
					\begin{tabular}{c c c c c c c }
						\toprule
						\multirow{2}{*}{Molecule} & \multirow{2}{*}{Type} & \multicolumn{2}{c}{DFTB} & \multicolumn{2}{c}{DFT} & \multirow{2}{*}{NN} \\ \cmidrule{3-6}
						& & FMO & $\Delta$-LC & $\Delta$-B3LYP & $\Delta$-$\omega$B97X & \\
						\midrule
						ANT & diag    & \num{3.2e+00} & \num{7.935e+02} & \num{2.1479215e+06} & \num{3.68104e+06} & \num{0.5} \\
						PEN & diag    & \num{1.02e+01} & \num{2.7527e+03} & - -             & - - & \num{0.9} \\
						\bottomrule
					\end{tabular}
				\end{center}
			\end{table}
			
			\begin{table}[ht]
				\caption{Comparison of timings (in ms) for the calculation of off-diagonal Hamiltonian elements (couplings) in anthracene and pentacene with DFTB and NN-models.}
				\label{tab:timings-offdiag}
				\begin{center}
					\begin{tabular}{c c c }
						\toprule
						Molecule & FMO-DFTB  & NN \\
						\midrule
						ANT      & \num{1.7} & \num{0.1} \\
						PEN      & \num{3.8} & \num{0.2} \\
						\bottomrule
					\end{tabular}
				\end{center}
			\end{table}
	
			The computational cost of diagonal Hamiltonian elements plus gradients in the FMO scheme is reduced by about one order of magnitude when turning from DFTB to neural network models.
			The same is true for off-diagonal Hamiltonian elements in the FMO approach.
			For the much more costly gradients calculated with LC-DFTB2, the respective neural network models outperform DFTB by more than three orders of magnitude.
			Additionally, the favorable $N^2$ scaling with the system size ($N$ being the number of atoms) of the NN models compared to the $N^3$ scaling of DFTB becomes apparent when comparing timings for anthracene (24 atoms) to pentacene (36 atoms).
			Switching to DFT gradients, the speedup for the NN models is about seven orders of magnitude.

\section{Conclusion}
\label{sec:conclusion}

\begin{acknowledgement}
	M.K., P.M.D. and M.E. gratefully acknowledge financial support by the German Research Foundation (DFG) through the Research Training Group 2450 “Tailored Scale-Bridging Approaches to Computational Nanoscience”.
	W.X. acknowledges support through the “Virtual Materials Design” (VIRTMAT) project.
	Additional support by the state of Baden-Württemberg through bwHPC and the German Research Foundation (DFG) through grant no INST 40/575-1 FUGG (JUSTUS 2 cluster) is appreciated.
\end{acknowledgement}



\bibliography{Manuscript.bib}

\providecommand{\latin}[1]{#1}
\makeatletter
\providecommand{\doi}
  {\begingroup\let\do\@makeother\dospecials
  \catcode`\{=1 \catcode`\}=2 \doi@aux}
\providecommand{\doi@aux}[1]{\endgroup\texttt{#1}}
\makeatother
\providecommand*\mcitethebibliography{\thebibliography}
\csname @ifundefined\endcsname{endmcitethebibliography}
  {\let\endmcitethebibliography\endthebibliography}{}
\begin{mcitethebibliography}{70}
\providecommand*\natexlab[1]{#1}
\providecommand*\mciteSetBstSublistMode[1]{}
\providecommand*\mciteSetBstMaxWidthForm[2]{}
\providecommand*\mciteBstWouldAddEndPuncttrue
  {\def\EndOfBibitem{\unskip.}}
\providecommand*\mciteBstWouldAddEndPunctfalse
  {\let\EndOfBibitem\relax}
\providecommand*\mciteSetBstMidEndSepPunct[3]{}
\providecommand*\mciteSetBstSublistLabelBeginEnd[3]{}
\providecommand*\EndOfBibitem{}
\mciteSetBstSublistMode{f}
\mciteSetBstMaxWidthForm{subitem}{(\alph{mcitesubitemcount})}
\mciteSetBstSublistLabelBeginEnd
  {\mcitemaxwidthsubitemform\space}
  {\relax}
  {\relax}

\bibitem[Armstrong \latin{et~al.}(2009)Armstrong, Wang, Alloway, Placencia,
  Ratcliff, and Brumbach]{armstrong2009_Organic}
Armstrong,~N.~R.; Wang,~W.; Alloway,~D.~M.; Placencia,~D.; Ratcliff,~E.;
  Brumbach,~M. Organic/Organic Heterojunctions: Organic Light Emitting Diodes
  and Organic Photovoltaic Devices. \emph{Macromol. Rapid Commun.}
  \textbf{2009}, \emph{30}, 717--731\relax
\mciteBstWouldAddEndPuncttrue
\mciteSetBstMidEndSepPunct{\mcitedefaultmidpunct}
{\mcitedefaultendpunct}{\mcitedefaultseppunct}\relax
\EndOfBibitem
\bibitem[Meerheim \latin{et~al.}(2009)Meerheim, Lussem, and
  Leo]{meerheim2009_Efficiency}
Meerheim,~R.; Lussem,~B.; Leo,~K. Efficiency and Stability of Pin Type Organic
  Light Emitting Diodes for Display and Lighting Applications. \emph{Proc.
  IEEE} \textbf{2009}, \emph{97}, 1606--1626\relax
\mciteBstWouldAddEndPuncttrue
\mciteSetBstMidEndSepPunct{\mcitedefaultmidpunct}
{\mcitedefaultendpunct}{\mcitedefaultseppunct}\relax
\EndOfBibitem
\bibitem[Kulkarni \latin{et~al.}(2004)Kulkarni, Tonzola, Babel, and
  Jenekhe]{kulkarni2004_Electron}
Kulkarni,~A.~P.; Tonzola,~C.~J.; Babel,~A.; Jenekhe,~S.~A. Electron Transport
  Materials for Organic Light-Emitting Diodes. \emph{Chem. Mater.}
  \textbf{2004}, \emph{16}, 4556--4573\relax
\mciteBstWouldAddEndPuncttrue
\mciteSetBstMidEndSepPunct{\mcitedefaultmidpunct}
{\mcitedefaultendpunct}{\mcitedefaultseppunct}\relax
\EndOfBibitem
\bibitem[Dimitrakopoulos and Malenfant(2002)Dimitrakopoulos, and
  Malenfant]{dimitrakopoulos2002_Organic}
Dimitrakopoulos,~C.~D.; Malenfant,~P.~R. Organic Thin Film Transistors for
  Large Area Electronics. \emph{Adv. Mater.} \textbf{2002}, \emph{14},
  99--117\relax
\mciteBstWouldAddEndPuncttrue
\mciteSetBstMidEndSepPunct{\mcitedefaultmidpunct}
{\mcitedefaultendpunct}{\mcitedefaultseppunct}\relax
\EndOfBibitem
\bibitem[Newman \latin{et~al.}(2004)Newman, Frisbie, {da Silva Filho},
  Br{\'e}das, Ewbank, and Mann]{newman2004_Introduction}
Newman,~C.~R.; Frisbie,~C.~D.; {da Silva Filho},~D.~A.; Br{\'e}das,~J.-L.;
  Ewbank,~P.~C.; Mann,~K.~R. Introduction to Organic Thin Film Transistors and
  Design of N-Channel Organic Semiconductors. \emph{Chem. Mater.}
  \textbf{2004}, \emph{16}, 4436--4451\relax
\mciteBstWouldAddEndPuncttrue
\mciteSetBstMidEndSepPunct{\mcitedefaultmidpunct}
{\mcitedefaultendpunct}{\mcitedefaultseppunct}\relax
\EndOfBibitem
\bibitem[Facchetti(2007)]{facchetti2007_Semiconductors}
Facchetti,~A. Semiconductors for Organic Transistors. \emph{Mater. Today}
  \textbf{2007}, \emph{10}, 28--37\relax
\mciteBstWouldAddEndPuncttrue
\mciteSetBstMidEndSepPunct{\mcitedefaultmidpunct}
{\mcitedefaultendpunct}{\mcitedefaultseppunct}\relax
\EndOfBibitem
\bibitem[Kippelen and Br{\'e}das(2009)Kippelen, and
  Br{\'e}das]{kippelen2009_Organic}
Kippelen,~B.; Br{\'e}das,~J.-L. Organic Photovoltaics. \emph{Energy Env. Sci}
  \textbf{2009}, \emph{2}, 251--261\relax
\mciteBstWouldAddEndPuncttrue
\mciteSetBstMidEndSepPunct{\mcitedefaultmidpunct}
{\mcitedefaultendpunct}{\mcitedefaultseppunct}\relax
\EndOfBibitem
\bibitem[Dennler \latin{et~al.}(2009)Dennler, Scharber, and
  Brabec]{dennler2009_Polymerfullerene}
Dennler,~G.; Scharber,~M.~C.; Brabec,~C.~J. Polymer-Fullerene
  Bulk-Heterojunction Solar Cells. \emph{Adv. Mater.} \textbf{2009}, \emph{21},
  1323--1338\relax
\mciteBstWouldAddEndPuncttrue
\mciteSetBstMidEndSepPunct{\mcitedefaultmidpunct}
{\mcitedefaultendpunct}{\mcitedefaultseppunct}\relax
\EndOfBibitem
\bibitem[Deibel and Dyakonov(2010)Deibel, and Dyakonov]{deibel2010_Polymer}
Deibel,~C.; Dyakonov,~V. Polymer\textendash Fullerene Bulk Heterojunction Solar
  Cells. \emph{Rep. Prog. Phys.} \textbf{2010}, \emph{73}, 096401\relax
\mciteBstWouldAddEndPuncttrue
\mciteSetBstMidEndSepPunct{\mcitedefaultmidpunct}
{\mcitedefaultendpunct}{\mcitedefaultseppunct}\relax
\EndOfBibitem
\bibitem[Dey \latin{et~al.}(2015)Dey, Singh, Das, and Iyer]{dey2015_Organic}
Dey,~A.; Singh,~A.; Das,~D.; Iyer,~P.~K. In \emph{Thin {{Film Structures}} in
  {{Energy Applications}}}; Babu Krishna~Moorthy,~S., Ed.; {Springer
  International Publishing}: {Cham}, 2015; pp 97--128\relax
\mciteBstWouldAddEndPuncttrue
\mciteSetBstMidEndSepPunct{\mcitedefaultmidpunct}
{\mcitedefaultendpunct}{\mcitedefaultseppunct}\relax
\EndOfBibitem
\bibitem[Oberhofer \latin{et~al.}(2017)Oberhofer, Reuter, and
  Blumberger]{oberhofer2017_Charge}
Oberhofer,~H.; Reuter,~K.; Blumberger,~J. Charge Transport in Molecular
  Materials: An Assessment of Computational Methods. \emph{Chem. Rev.}
  \textbf{2017}, \emph{117}, 10319--10357\relax
\mciteBstWouldAddEndPuncttrue
\mciteSetBstMidEndSepPunct{\mcitedefaultmidpunct}
{\mcitedefaultendpunct}{\mcitedefaultseppunct}\relax
\EndOfBibitem
\bibitem[Kuba{\v r} \latin{et~al.}(2013)Kuba{\v r}, Guti{\'e}rrez,
  Kleinekath{\"o}fer, Cuniberti, and Elstner]{kubar2013_Modeling}
Kuba{\v r},~T.; Guti{\'e}rrez,~R.; Kleinekath{\"o}fer,~U.; Cuniberti,~G.;
  Elstner,~M. Modeling Charge Transport in {{DNA}} Using Multi-Scale Methods.
  \emph{Phys. Status Solidi B} \textbf{2013}, \emph{250}, 2277--2287\relax
\mciteBstWouldAddEndPuncttrue
\mciteSetBstMidEndSepPunct{\mcitedefaultmidpunct}
{\mcitedefaultendpunct}{\mcitedefaultseppunct}\relax
\EndOfBibitem
\bibitem[Heck \latin{et~al.}(2015)Heck, Kranz, Kuba{\v r}, and
  Elstner]{heck2015_Multiscale}
Heck,~A.; Kranz,~J.~J.; Kuba{\v r},~T.; Elstner,~M. Multi-Scale Approach to
  Non-Adiabatic Charge Transport in High-Mobility Organic Semiconductors.
  \emph{J. Chem. Theory Comput.} \textbf{2015}, \emph{11}, 5068--5082\relax
\mciteBstWouldAddEndPuncttrue
\mciteSetBstMidEndSepPunct{\mcitedefaultmidpunct}
{\mcitedefaultendpunct}{\mcitedefaultseppunct}\relax
\EndOfBibitem
\bibitem[Kranz and Elstner(2016)Kranz, and Elstner]{kranz2016_Simulation}
Kranz,~J.~J.; Elstner,~M. Simulation of {{Singlet Exciton Diffusion}} in {{Bulk
  Organic Materials}}. \emph{J. Chem. Theory Comput.} \textbf{2016}, \emph{12},
  4209--4221\relax
\mciteBstWouldAddEndPuncttrue
\mciteSetBstMidEndSepPunct{\mcitedefaultmidpunct}
{\mcitedefaultendpunct}{\mcitedefaultseppunct}\relax
\EndOfBibitem
\bibitem[Giannini \latin{et~al.}(2019)Giannini, Carof, Ellis, Yang, Ziogos,
  Ghosh, and Blumberger]{giannini2019_Quantum}
Giannini,~S.; Carof,~A.; Ellis,~M.; Yang,~H.; Ziogos,~O.~G.; Ghosh,~S.;
  Blumberger,~J. Quantum Localization and Delocalization of Charge Carriers in
  Organic Semiconducting Crystals. \emph{Nat. Commun.} \textbf{2019},
  \emph{10}, 1--12\relax
\mciteBstWouldAddEndPuncttrue
\mciteSetBstMidEndSepPunct{\mcitedefaultmidpunct}
{\mcitedefaultendpunct}{\mcitedefaultseppunct}\relax
\EndOfBibitem
\bibitem[Lederer \latin{et~al.}(2019)Lederer, Kaiser, Mattoni, and
  Gagliardi]{lederer2019_Machine}
Lederer,~J.; Kaiser,~W.; Mattoni,~A.; Gagliardi,~A. Machine
  {{Learning}}\textendash{{Based Charge Transport Computation}} for
  {{Pentacene}}. \emph{Adv. Theory Simul.} \textbf{2019}, \emph{2},
  1800136--1800136\relax
\mciteBstWouldAddEndPuncttrue
\mciteSetBstMidEndSepPunct{\mcitedefaultmidpunct}
{\mcitedefaultendpunct}{\mcitedefaultseppunct}\relax
\EndOfBibitem
\bibitem[Wang \latin{et~al.}(2019)Wang, Braza, Claudio, Nellas, and
  Hsu]{wang2019_Machine}
Wang,~C.-I.; Braza,~M. K.~E.; Claudio,~G.~C.; Nellas,~R.~B.; Hsu,~C.-P. Machine
  {{Learning}} for {{Predicting Electron Transfer Coupling}}. \emph{J. Phys.
  Chem. A} \textbf{2019}, \emph{123}, 7792--7802\relax
\mciteBstWouldAddEndPuncttrue
\mciteSetBstMidEndSepPunct{\mcitedefaultmidpunct}
{\mcitedefaultendpunct}{\mcitedefaultseppunct}\relax
\EndOfBibitem
\bibitem[Kr{\"a}mer \latin{et~al.}(2020)Kr{\"a}mer, Dohmen, Xie, Holub,
  Christensen, and Elstner]{kramer2020_Charge}
Kr{\"a}mer,~M.; Dohmen,~P.~M.; Xie,~W.; Holub,~D.; Christensen,~A.~S.;
  Elstner,~M. Charge and {{Exciton Transfer Simulations Using
  Machine}}-{{Learned Hamiltonians}}. \emph{J. Chem. Theory Comput.}
  \textbf{2020}, \emph{16}, 4061--4070\relax
\mciteBstWouldAddEndPuncttrue
\mciteSetBstMidEndSepPunct{\mcitedefaultmidpunct}
{\mcitedefaultendpunct}{\mcitedefaultseppunct}\relax
\EndOfBibitem
\bibitem[Musil \latin{et~al.}(2018)Musil, De, Yang, Campbell, Day, and
  Ceriotti]{musil2018_Machine}
Musil,~F.; De,~S.; Yang,~J.; Campbell,~J.~E.; Day,~G.~M.; Ceriotti,~M. Machine
  Learning for the Structure\textendash Energy\textendash Property Landscapes
  of Molecular Crystals. \emph{Chem. Sci.} \textbf{2018}, \emph{9},
  1289--1300\relax
\mciteBstWouldAddEndPuncttrue
\mciteSetBstMidEndSepPunct{\mcitedefaultmidpunct}
{\mcitedefaultendpunct}{\mcitedefaultseppunct}\relax
\EndOfBibitem
\bibitem[{\c C}aylak \latin{et~al.}(2019){\c C}aylak, Yaman, and
  Baumeier]{caylak2019_Evolutionary}
{\c C}aylak,~O.; Yaman,~A.; Baumeier,~B. Evolutionary {{Approach}} to
  {{Constructing}} a {{Deep Feedforward Neural Network}} for {{Prediction}} of
  {{Electronic Coupling Elements}} in {{Molecular Materials}}. \emph{J. Chem.
  Theory Comput.} \textbf{2019}, \emph{15}, 1777--1784\relax
\mciteBstWouldAddEndPuncttrue
\mciteSetBstMidEndSepPunct{\mcitedefaultmidpunct}
{\mcitedefaultendpunct}{\mcitedefaultseppunct}\relax
\EndOfBibitem
\bibitem[H{\"a}se \latin{et~al.}(2016)H{\"a}se, Valleau, {Pyzer-Knapp}, and
  {Aspuru-Guzik}]{hase2016_Machine}
H{\"a}se,~F.; Valleau,~S.; {Pyzer-Knapp},~E.; {Aspuru-Guzik},~A. Machine
  Learning Exciton Dynamics. \emph{Chem. Sci.} \textbf{2016}, \emph{7},
  5139--5147\relax
\mciteBstWouldAddEndPuncttrue
\mciteSetBstMidEndSepPunct{\mcitedefaultmidpunct}
{\mcitedefaultendpunct}{\mcitedefaultseppunct}\relax
\EndOfBibitem
\bibitem[Li \latin{et~al.}(2021)Li, Reiser, Boswell, Eberhard, Burns,
  Friederich, and Lopez]{li2021_Automatic}
Li,~J.; Reiser,~P.; Boswell,~B.~R.; Eberhard,~A.; Burns,~N.~Z.; Friederich,~P.;
  Lopez,~S.~A. Automatic Discovery of Photoisomerization Mechanisms with
  Nanosecond Machine Learning Photodynamics Simulations. \emph{Chem. Sci.}
  \textbf{2021}, \relax
\mciteBstWouldAddEndPunctfalse
\mciteSetBstMidEndSepPunct{\mcitedefaultmidpunct}
{}{\mcitedefaultseppunct}\relax
\EndOfBibitem
\bibitem[Westermayr \latin{et~al.}(2019)Westermayr, Gastegger, Menger, Mai,
  Gonz{\'a}lez, and Marquetand]{westermayr2019_Machine}
Westermayr,~J.; Gastegger,~M.; Menger,~M. F. S.~J.; Mai,~S.; Gonz{\'a}lez,~L.;
  Marquetand,~P. Machine Learning Enables Long Time Scale Molecular
  Photodynamics Simulations. \emph{Chem. Sci.} \textbf{2019}, \emph{10},
  8100--8107\relax
\mciteBstWouldAddEndPuncttrue
\mciteSetBstMidEndSepPunct{\mcitedefaultmidpunct}
{\mcitedefaultendpunct}{\mcitedefaultseppunct}\relax
\EndOfBibitem
\bibitem[Westermayr and Marquetand(2020)Westermayr, and
  Marquetand]{westermayr2020_Machinea}
Westermayr,~J.; Marquetand,~P. Machine Learning and Excited-State Molecular
  Dynamics. \emph{Mach. Learn.: Sci. Technol.} \textbf{2020}, \emph{1},
  043001\relax
\mciteBstWouldAddEndPuncttrue
\mciteSetBstMidEndSepPunct{\mcitedefaultmidpunct}
{\mcitedefaultendpunct}{\mcitedefaultseppunct}\relax
\EndOfBibitem
\bibitem[Westermayr \latin{et~al.}(2020)Westermayr, Gastegger, and
  Marquetand]{westermayr2020_Combining}
Westermayr,~J.; Gastegger,~M.; Marquetand,~P. Combining {{SchNet}} and
  {{SHARC}}: The {{SchNarc Machine Learning Approach}} for {{Excited}}-{{State
  Dynamics}}. \emph{J. Phys. Chem. Lett.} \textbf{2020}, \emph{11},
  3828--3834\relax
\mciteBstWouldAddEndPuncttrue
\mciteSetBstMidEndSepPunct{\mcitedefaultmidpunct}
{\mcitedefaultendpunct}{\mcitedefaultseppunct}\relax
\EndOfBibitem
\bibitem[Porezag \latin{et~al.}(1995)Porezag, Frauenheim, K{\"o}hler, Seifert,
  and Kaschner]{porezag1995_Construction}
Porezag,~D.; Frauenheim,~T.; K{\"o}hler,~T.; Seifert,~G.; Kaschner,~R.
  Construction of Tight-Binding-like Potentials on the Basis of
  Density-Functional Theory: Application to Carbon. \emph{Phys. Rev. B}
  \textbf{1995}, \emph{51}, 12947\relax
\mciteBstWouldAddEndPuncttrue
\mciteSetBstMidEndSepPunct{\mcitedefaultmidpunct}
{\mcitedefaultendpunct}{\mcitedefaultseppunct}\relax
\EndOfBibitem
\bibitem[Seifert \latin{et~al.}(1996)Seifert, Porezag, and
  Frauenheim]{seifert1996_Calculations}
Seifert,~G.; Porezag,~D.; Frauenheim,~T. Calculations of Molecules, Clusters,
  and Solids with a Simplified {{LCAO}}-{{DFT}}-{{LDA}} Scheme. \emph{Int. J.
  Quantum Chem.} \textbf{1996}, \emph{58}, 185--192\relax
\mciteBstWouldAddEndPuncttrue
\mciteSetBstMidEndSepPunct{\mcitedefaultmidpunct}
{\mcitedefaultendpunct}{\mcitedefaultseppunct}\relax
\EndOfBibitem
\bibitem[Elstner \latin{et~al.}(1998)Elstner, Porezag, Jungnickel, Elsner,
  Haugk, Frauenheim, Suhai, and Seifert]{elstner1998_Selfconsistentcharge}
Elstner,~M.; Porezag,~D.; Jungnickel,~G.; Elsner,~J.; Haugk,~M.;
  Frauenheim,~T.; Suhai,~S.; Seifert,~G. Self-Consistent-Charge
  Density-Functional Tight-Binding Method for Simulations of Complex Materials
  Properties. \emph{Phys. Rev. B} \textbf{1998}, \emph{58}, 7260\relax
\mciteBstWouldAddEndPuncttrue
\mciteSetBstMidEndSepPunct{\mcitedefaultmidpunct}
{\mcitedefaultendpunct}{\mcitedefaultseppunct}\relax
\EndOfBibitem
\bibitem[Niehaus \latin{et~al.}(2001)Niehaus, Suhai, Della~Sala, Lugli,
  Elstner, Seifert, and Frauenheim]{niehaus2001_Tightbinding}
Niehaus,~T.~A.; Suhai,~S.; Della~Sala,~F.; Lugli,~P.; Elstner,~M.; Seifert,~G.;
  Frauenheim,~T. Tight-Binding Approach to Time-Dependent Density-Functional
  Response Theory. \emph{Phys. Rev. B} \textbf{2001}, \emph{63}, 085108\relax
\mciteBstWouldAddEndPuncttrue
\mciteSetBstMidEndSepPunct{\mcitedefaultmidpunct}
{\mcitedefaultendpunct}{\mcitedefaultseppunct}\relax
\EndOfBibitem
\bibitem[Kranz \latin{et~al.}(2017)Kranz, Elstner, Aradi, Frauenheim, Lutsker,
  Garcia, and Niehaus]{kranz2017_Timedependent}
Kranz,~J.~J.; Elstner,~M.; Aradi,~B.; Frauenheim,~T.; Lutsker,~V.;
  Garcia,~A.~D.; Niehaus,~T.~A. Time-Dependent Extension of the Long-Range
  Corrected Density Functional Based Tight-Binding Method. \emph{J. Chem.
  Theory Comput.} \textbf{2017}, \emph{13}, 1737--1747\relax
\mciteBstWouldAddEndPuncttrue
\mciteSetBstMidEndSepPunct{\mcitedefaultmidpunct}
{\mcitedefaultendpunct}{\mcitedefaultseppunct}\relax
\EndOfBibitem
\bibitem[Aradi \latin{et~al.}(2007)Aradi, Hourahine, and
  Frauenheim]{aradi2007_DFTB}
Aradi,~B.; Hourahine,~B.; Frauenheim,~T. {{DFTB}}+, a Sparse Matrix-Based
  Implementation of the {{DFTB}} Method. \emph{J. Phys. Chem A} \textbf{2007},
  \emph{111}, 5678--5684\relax
\mciteBstWouldAddEndPuncttrue
\mciteSetBstMidEndSepPunct{\mcitedefaultmidpunct}
{\mcitedefaultendpunct}{\mcitedefaultseppunct}\relax
\EndOfBibitem
\bibitem[Hourahine \latin{et~al.}(2020)Hourahine, Aradi, Blum, Bonaf{\'e},
  Buccheri, Camacho, Cevallos, Deshaye, Dumitric{\u a}, Dominguez, Ehlert,
  Elstner, {van der Heide}, Hermann, Irle, Kranz, K{\"o}hler, Kowalczyk,
  Kuba{\v r}, Lee, Lutsker, Maurer, Min, Mitchell, Negre, Niehaus, Niklasson,
  Page, Pecchia, Penazzi, Persson, {\v R}ez{\'a}{\v c}, S{\'a}nchez, Sternberg,
  St{\"o}hr, Stuckenberg, Tkatchenko, Yu, and Frauenheim]{hourahine2020_DFTB}
Hourahine,~B.; Aradi,~B.; Blum,~V.; Bonaf{\'e},~F.; Buccheri,~A.; Camacho,~C.;
  Cevallos,~C.; Deshaye,~M.~Y.; Dumitric{\u a},~T.; Dominguez,~A.; Ehlert,~S.;
  Elstner,~M.; {van der Heide},~T.; Hermann,~J.; Irle,~S.; Kranz,~J.~J.;
  K{\"o}hler,~C.; Kowalczyk,~T.; Kuba{\v r},~T.; Lee,~I.~S.; Lutsker,~V.;
  Maurer,~R.~J.; Min,~S.~K.; Mitchell,~I.; Negre,~C.; Niehaus,~T.~A.;
  Niklasson,~A. M.~N.; Page,~A.~J.; Pecchia,~A.; Penazzi,~G.; Persson,~M.~P.;
  {\v R}ez{\'a}{\v c},~J.; S{\'a}nchez,~C.~G.; Sternberg,~M.; St{\"o}hr,~M.;
  Stuckenberg,~F.; Tkatchenko,~A.; Yu,~V. W.-z.; Frauenheim,~T. {{DFTB}}+, a
  Software Package for Efficient Approximate Density Functional Theory Based
  Atomistic Simulations. \emph{J. Chem. Phys.} \textbf{2020}, \emph{152},
  124101\relax
\mciteBstWouldAddEndPuncttrue
\mciteSetBstMidEndSepPunct{\mcitedefaultmidpunct}
{\mcitedefaultendpunct}{\mcitedefaultseppunct}\relax
\EndOfBibitem
\bibitem[Christensen \latin{et~al.}(2019)Christensen, Faber, and {von
  Lilienfeld}]{christensen2019_Operators}
Christensen,~A.~S.; Faber,~F.~A.; {von Lilienfeld},~O.~A. Operators in Quantum
  Machine Learning: Response Properties in Chemical Space. \emph{J. Chem.
  Phys.} \textbf{2019}, \emph{150}, 064105\relax
\mciteBstWouldAddEndPuncttrue
\mciteSetBstMidEndSepPunct{\mcitedefaultmidpunct}
{\mcitedefaultendpunct}{\mcitedefaultseppunct}\relax
\EndOfBibitem
\bibitem[Kuba{\v r} and Elstner(2013)Kuba{\v r}, and Elstner]{kubar2013_hybrid}
Kuba{\v r},~T.; Elstner,~M. A Hybrid Approach to Simulation of Electron
  Transfer in Complex Molecular Systems. \emph{J R Soc Interface}
  \textbf{2013}, \emph{10}, 20130415\relax
\mciteBstWouldAddEndPuncttrue
\mciteSetBstMidEndSepPunct{\mcitedefaultmidpunct}
{\mcitedefaultendpunct}{\mcitedefaultseppunct}\relax
\EndOfBibitem
\bibitem[Kitaura \latin{et~al.}(1999)Kitaura, Ikeo, Asada, Nakano, and
  Uebayasi]{kitaura1999_Fragment}
Kitaura,~K.; Ikeo,~E.; Asada,~T.; Nakano,~T.; Uebayasi,~M. Fragment Molecular
  Orbital Method: An Approximate Computational Method for Large Molecules.
  \emph{Chem. Phys. Lett.} \textbf{1999}, \emph{313}, 701--706\relax
\mciteBstWouldAddEndPuncttrue
\mciteSetBstMidEndSepPunct{\mcitedefaultmidpunct}
{\mcitedefaultendpunct}{\mcitedefaultseppunct}\relax
\EndOfBibitem
\bibitem[Kuba{\v r} \latin{et~al.}(2008)Kuba{\v r}, Woiczikowski, Cuniberti,
  and Elstner]{kubar2008_Efficient}
Kuba{\v r},~T.; Woiczikowski,~P.~B.; Cuniberti,~G.; Elstner,~M. Efficient
  Calculation of Charge-Transfer Matrix Elements for Hole Transfer in {{DNA}}.
  \emph{J. Phys. Chem. B} \textbf{2008}, \emph{112}, 7937--7947\relax
\mciteBstWouldAddEndPuncttrue
\mciteSetBstMidEndSepPunct{\mcitedefaultmidpunct}
{\mcitedefaultendpunct}{\mcitedefaultseppunct}\relax
\EndOfBibitem
\bibitem[Ehrenfest(1927)]{ehrenfest1927_Bemerkung}
Ehrenfest,~P. Bemerkung \"Uber Die Angen\"aherte {{G\"ultigkeit}} Der
  Klassischen {{Mechanik}} Innerhalb Der {{Quantenmechanik}}. \emph{Z. Phys.}
  \textbf{1927}, \emph{45}, 455--457\relax
\mciteBstWouldAddEndPuncttrue
\mciteSetBstMidEndSepPunct{\mcitedefaultmidpunct}
{\mcitedefaultendpunct}{\mcitedefaultseppunct}\relax
\EndOfBibitem
\bibitem[McLachlan(1964)]{mclachlan1964_variational}
McLachlan,~A. A Variational Solution of the Time-Dependent {{Schrodinger}}
  Equation. \emph{Mol. Phys.} \textbf{1964}, \emph{8}, 39--44\relax
\mciteBstWouldAddEndPuncttrue
\mciteSetBstMidEndSepPunct{\mcitedefaultmidpunct}
{\mcitedefaultendpunct}{\mcitedefaultseppunct}\relax
\EndOfBibitem
\bibitem[Tully(1990)]{tully1990_Molecular}
Tully,~J.~C. Molecular Dynamics with Electronic Transitions. \emph{J. Chem.
  Phys.} \textbf{1990}, \emph{93}, 1061--1071\relax
\mciteBstWouldAddEndPuncttrue
\mciteSetBstMidEndSepPunct{\mcitedefaultmidpunct}
{\mcitedefaultendpunct}{\mcitedefaultseppunct}\relax
\EndOfBibitem
\bibitem[Carof \latin{et~al.}(2017)Carof, Giannini, and
  Blumberger]{carof2017_Detailed}
Carof,~A.; Giannini,~S.; Blumberger,~J. Detailed Balance, Internal Consistency,
  and Energy Conservation in Fragment Orbital-Based Surface Hopping. \emph{J.
  Chem. Phys.} \textbf{2017}, \emph{147}, 214113\relax
\mciteBstWouldAddEndPuncttrue
\mciteSetBstMidEndSepPunct{\mcitedefaultmidpunct}
{\mcitedefaultendpunct}{\mcitedefaultseppunct}\relax
\EndOfBibitem
\bibitem[Akimov and Prezhdo(2013)Akimov, and Prezhdo]{akimov2013_PYXAID}
Akimov,~A.~V.; Prezhdo,~O.~V. The {{PYXAID}} Program for Non-Adiabatic
  Molecular Dynamics in Condensed Matter Systems. \emph{J. Chem. Theory
  Comput.} \textbf{2013}, \emph{9}, 4959--4972\relax
\mciteBstWouldAddEndPuncttrue
\mciteSetBstMidEndSepPunct{\mcitedefaultmidpunct}
{\mcitedefaultendpunct}{\mcitedefaultseppunct}\relax
\EndOfBibitem
\bibitem[Mason(1964)]{mason1964_crystallography}
Mason,~R. The Crystallography of Anthracene at 95\$\^\textbackslash circ\${{K}}
  and 290\$\^\textbackslash circ\${{K}}. \emph{Acta Crystallogr.}
  \textbf{1964}, \emph{17}, 547--555\relax
\mciteBstWouldAddEndPuncttrue
\mciteSetBstMidEndSepPunct{\mcitedefaultmidpunct}
{\mcitedefaultendpunct}{\mcitedefaultseppunct}\relax
\EndOfBibitem
\bibitem[Wang \latin{et~al.}(2004)Wang, Wolf, Caldwell, Kollman, and
  Case]{wang2004_Development}
Wang,~J.; Wolf,~R.~M.; Caldwell,~J.~W.; Kollman,~P.~A.; Case,~D.~A. Development
  and Testing of a General Amber Force Field. \emph{J. Comput. Chem.}
  \textbf{2004}, \emph{25}, 1157--1174\relax
\mciteBstWouldAddEndPuncttrue
\mciteSetBstMidEndSepPunct{\mcitedefaultmidpunct}
{\mcitedefaultendpunct}{\mcitedefaultseppunct}\relax
\EndOfBibitem
\bibitem[Wang \latin{et~al.}(2006)Wang, Wang, Kollman, and
  Case]{wang2006_Automatic}
Wang,~J.; Wang,~W.; Kollman,~P.~A.; Case,~D.~A. Automatic Atom Type and Bond
  Type Perception in Molecular Mechanical Calculations. \emph{J. Mol. Graph.
  Model.} \textbf{2006}, \emph{25}, 247--260\relax
\mciteBstWouldAddEndPuncttrue
\mciteSetBstMidEndSepPunct{\mcitedefaultmidpunct}
{\mcitedefaultendpunct}{\mcitedefaultseppunct}\relax
\EndOfBibitem
\bibitem[Singh and Kollman(1984)Singh, and Kollman]{singh1984_approach}
Singh,~U.~C.; Kollman,~P.~A. An Approach to Computing Electrostatic Charges for
  Molecules. \emph{J. Comput. Chem.} \textbf{1984}, \emph{5}, 129--145\relax
\mciteBstWouldAddEndPuncttrue
\mciteSetBstMidEndSepPunct{\mcitedefaultmidpunct}
{\mcitedefaultendpunct}{\mcitedefaultseppunct}\relax
\EndOfBibitem
\bibitem[Besler \latin{et~al.}(1990)Besler, Merz, and
  Kollman]{besler1990_Atomic}
Besler,~B.~H.; Merz,~K.~M.; Kollman,~P.~A. Atomic Charges Derived from
  Semiempirical Methods. \emph{J. Comput. Chem.} \textbf{1990}, \emph{11},
  431--439\relax
\mciteBstWouldAddEndPuncttrue
\mciteSetBstMidEndSepPunct{\mcitedefaultmidpunct}
{\mcitedefaultendpunct}{\mcitedefaultseppunct}\relax
\EndOfBibitem
\bibitem[Petersson \latin{et~al.}(1988)Petersson, Bennett, Tensfeldt,
  {Al-Laham}, Shirley, and Mantzaris]{petersson1988_complete}
Petersson,~G.~A.; Bennett,~A.; Tensfeldt,~T.~G.; {Al-Laham},~M.~A.;
  Shirley,~W.~A.; Mantzaris,~J. A Complete Basis Set Model Chemistry. {{I}}.
  {{The}} Total Energies of Closed-Shell Atoms and Hydrides of the First-Row
  Elements. \emph{J. Chem. Phys.} \textbf{1988}, \emph{89}, 2193--2218\relax
\mciteBstWouldAddEndPuncttrue
\mciteSetBstMidEndSepPunct{\mcitedefaultmidpunct}
{\mcitedefaultendpunct}{\mcitedefaultseppunct}\relax
\EndOfBibitem
\bibitem[Petersson and {Al-Laham}(1991)Petersson, and
  {Al-Laham}]{petersson1991_complete}
Petersson,~G.~A.; {Al-Laham},~M.~A. A Complete Basis Set Model Chemistry.
  {{II}}. {{Open}}-Shell Systems and the Total Energies of the First-Row Atoms.
  \emph{J. Chem. Phys.} \textbf{1991}, \emph{94}, 6081--6090\relax
\mciteBstWouldAddEndPuncttrue
\mciteSetBstMidEndSepPunct{\mcitedefaultmidpunct}
{\mcitedefaultendpunct}{\mcitedefaultseppunct}\relax
\EndOfBibitem
\bibitem[Frisch \latin{et~al.}(2009)Frisch, Trucks, Schlegel, Scuseria, Robb,
  Cheeseman, Scalmani, Barone, Mennucci, Petersson, Nakatsuji, Caricato, Li,
  Hratchian, Izmaylov, Bloino, Zheng, Sonnenberg, Hada, Ehara, Toyota, Fukuda,
  Hasegawa, Ishida, Nakajima, Honda, Kitao, Nakai, Vreven, Montgomery~Jr.,
  Peralta, Ogliaro, Bearpark, Heyd, Brothers, Kudin, Staroverov, Kobayashi,
  Normand, Raghavachari, Rendell, Burant, Iyengar, Tomasi, Cossi, Rega, Millam,
  Klene, Knox, Cross, Bakken, Adamo, Jaramillo, Gomperts, Stratmann, Yazyev,
  Austin, Cammi, Pomelli, Ochterski, Martin, Morokuma, Zakrzewski, Voth,
  Salvador, Dannenberg, Dapprich, Daniels, Farkas, Foresman, Ortiz, Cioslowski,
  and Fox]{frisch2009_Gaussian}
Frisch,~M.~J.; Trucks,~G.~W.; Schlegel,~H.~B.; Scuseria,~G.~E.; Robb,~M.~A.;
  Cheeseman,~J.~R.; Scalmani,~G.; Barone,~V.; Mennucci,~B.; Petersson,~G.~A.;
  Nakatsuji,~H.; Caricato,~M.; Li,~X.; Hratchian,~H.~P.; Izmaylov,~A.~F.;
  Bloino,~J.; Zheng,~G.; Sonnenberg,~J.~L.; Hada,~M.; Ehara,~M.; Toyota,~K.;
  Fukuda,~R.; Hasegawa,~J.; Ishida,~M.; Nakajima,~T.; Honda,~Y.; Kitao,~O.;
  Nakai,~H.; Vreven,~T.; Montgomery~Jr.,~J.~A.; Peralta,~J.~E.; Ogliaro,~F.;
  Bearpark,~M.~J.; Heyd,~J.; Brothers,~E.~N.; Kudin,~K.~N.; Staroverov,~V.~N.;
  Kobayashi,~R.; Normand,~J.; Raghavachari,~K.; Rendell,~A.~P.; Burant,~J.~C.;
  Iyengar,~S.~S.; Tomasi,~J.; Cossi,~M.; Rega,~N.; Millam,~N.~J.; Klene,~M.;
  Knox,~J.~E.; Cross,~J.~B.; Bakken,~V.; Adamo,~C.; Jaramillo,~J.;
  Gomperts,~R.; Stratmann,~R.~E.; Yazyev,~O.; Austin,~A.~J.; Cammi,~R.;
  Pomelli,~C.; Ochterski,~J.~W.; Martin,~R.~L.; Morokuma,~K.;
  Zakrzewski,~V.~G.; Voth,~G.~A.; Salvador,~P.; Dannenberg,~J.~J.;
  Dapprich,~S.; Daniels,~A.~D.; Farkas,~{\"O}.; Foresman,~J.~B.; Ortiz,~J.~V.;
  Cioslowski,~J.; Fox,~D.~J. \emph{Gaussian 09}; {Gaussian, Inc.}:
  {Wallingford, CT, USA}, 2009\relax
\mciteBstWouldAddEndPuncttrue
\mciteSetBstMidEndSepPunct{\mcitedefaultmidpunct}
{\mcitedefaultendpunct}{\mcitedefaultseppunct}\relax
\EndOfBibitem
\bibitem[Evans and Holian(1985)Evans, and Holian]{evans1985_Nose}
Evans,~D.~J.; Holian,~B.~L. The {{Nose}}\textendash{{Hoover}} Thermostat.
  \emph{J. Chem. Phys.} \textbf{1985}, \emph{83}, 4069\relax
\mciteBstWouldAddEndPuncttrue
\mciteSetBstMidEndSepPunct{\mcitedefaultmidpunct}
{\mcitedefaultendpunct}{\mcitedefaultseppunct}\relax
\EndOfBibitem
\bibitem[Berendsen \latin{et~al.}(1995)Berendsen, {van der Spoel}, and {van
  Drunen}]{berendsen1995_GROMACS}
Berendsen,~H. J.~C.; {van der Spoel},~D.; {van Drunen},~R. {{GROMACS}}: A
  Message-Passing Parallel Molecular Dynamics Implementation. \emph{Comput.
  Phys. Commun.} \textbf{1995}, \emph{91}, 43--56\relax
\mciteBstWouldAddEndPuncttrue
\mciteSetBstMidEndSepPunct{\mcitedefaultmidpunct}
{\mcitedefaultendpunct}{\mcitedefaultseppunct}\relax
\EndOfBibitem
\bibitem[Abraham \latin{et~al.}(2015)Abraham, Murtola, Schulz, P{\'a}ll, Smith,
  Hess, and Lindahl]{abraham2015_GROMACS}
Abraham,~M.~J.; Murtola,~T.; Schulz,~R.; P{\'a}ll,~S.; Smith,~J.~C.; Hess,~B.;
  Lindahl,~E. {{GROMACS}}: High Performance Molecular Simulations through
  Multi-Level Parallelism from Laptops to Supercomputers. \emph{SoftwareX}
  \textbf{2015}, \emph{1--2}, 19--25\relax
\mciteBstWouldAddEndPuncttrue
\mciteSetBstMidEndSepPunct{\mcitedefaultmidpunct}
{\mcitedefaultendpunct}{\mcitedefaultseppunct}\relax
\EndOfBibitem
\bibitem[Niehaus and Della~Sala(2012)Niehaus, and
  Della~Sala]{niehaus2012_Range}
Niehaus,~T.~A.; Della~Sala,~F. Range Separated Functionals in the Density
  Functional Based Tight-Binding Method: Formalism. \emph{Phys. Status Solidi
  B} \textbf{2012}, \emph{249}, 237--244\relax
\mciteBstWouldAddEndPuncttrue
\mciteSetBstMidEndSepPunct{\mcitedefaultmidpunct}
{\mcitedefaultendpunct}{\mcitedefaultseppunct}\relax
\EndOfBibitem
\bibitem[Becke(1993)]{becke1993becke}
Becke,~A.~D. Becke’s three parameter hybrid method using the LYP correlation
  functional. \emph{J. Chem. Phys.} \textbf{1993}, \emph{98}, 5648--5652\relax
\mciteBstWouldAddEndPuncttrue
\mciteSetBstMidEndSepPunct{\mcitedefaultmidpunct}
{\mcitedefaultendpunct}{\mcitedefaultseppunct}\relax
\EndOfBibitem
\bibitem[Chai and Head-Gordon(2008)Chai, and Head-Gordon]{chai2008systematic}
Chai,~J.-D.; Head-Gordon,~M. Systematic optimization of long-range corrected
  hybrid density functionals. \emph{J. Chem. Phys.} \textbf{2008}, \emph{128},
  084106\relax
\mciteBstWouldAddEndPuncttrue
\mciteSetBstMidEndSepPunct{\mcitedefaultmidpunct}
{\mcitedefaultendpunct}{\mcitedefaultseppunct}\relax
\EndOfBibitem
\bibitem[Weigend and Ahlrichs(2005)Weigend, and Ahlrichs]{weigend2005balanced}
Weigend,~F.; Ahlrichs,~R. Balanced basis sets of split valence, triple zeta
  valence and quadruple zeta valence quality for H to Rn: Design and assessment
  of accuracy. \emph{Phys. Chem. Chem. Phys.} \textbf{2005}, \emph{7},
  3297--3305\relax
\mciteBstWouldAddEndPuncttrue
\mciteSetBstMidEndSepPunct{\mcitedefaultmidpunct}
{\mcitedefaultendpunct}{\mcitedefaultseppunct}\relax
\EndOfBibitem
\bibitem[Weigend(2006)]{weigend2006accurate}
Weigend,~F. Accurate Coulomb-fitting basis sets for H to Rn. \emph{Phys. Chem.
  Chem. Phys.} \textbf{2006}, \emph{8}, 1057--1065\relax
\mciteBstWouldAddEndPuncttrue
\mciteSetBstMidEndSepPunct{\mcitedefaultmidpunct}
{\mcitedefaultendpunct}{\mcitedefaultseppunct}\relax
\EndOfBibitem
\bibitem[Neese(2012)]{neese2012orca}
Neese,~F. The ORCA program system. \emph{Wiley Interdiscip. Rev. Comput. Mol.
  Sci.} \textbf{2012}, \emph{2}, 73--78\relax
\mciteBstWouldAddEndPuncttrue
\mciteSetBstMidEndSepPunct{\mcitedefaultmidpunct}
{\mcitedefaultendpunct}{\mcitedefaultseppunct}\relax
\EndOfBibitem
\bibitem[Neese(2018)]{neese2018software}
Neese,~F. Software update: the ORCA program system, version 4.0. \emph{Wiley
  Interdiscip. Rev. Comput. Mol. Sci.} \textbf{2018}, \emph{8}, e1327\relax
\mciteBstWouldAddEndPuncttrue
\mciteSetBstMidEndSepPunct{\mcitedefaultmidpunct}
{\mcitedefaultendpunct}{\mcitedefaultseppunct}\relax
\EndOfBibitem
\bibitem[Zhao \latin{et~al.}(2018)Zhao, Liu, Li, and Luo]{zhao2018_novel}
Zhao,~H.; Liu,~F.; Li,~L.; Luo,~C. A Novel Softplus Linear Unit for Deep
  Convolutional Neural Networks. \emph{Appl Intell} \textbf{2018}, \emph{48},
  1707--1720\relax
\mciteBstWouldAddEndPuncttrue
\mciteSetBstMidEndSepPunct{\mcitedefaultmidpunct}
{\mcitedefaultendpunct}{\mcitedefaultseppunct}\relax
\EndOfBibitem
\bibitem[{Mart\'in Abadi} \latin{et~al.}(2015){Mart\'in Abadi}, {Ashish
  Agarwal}, {Paul Barham}, {Eugene Brevdo}, {Zhifeng Chen}, {Craig Citro},
  {Greg S. Corrado}, {Andy Davis}, {Jeffrey Dean}, {Matthieu Devin}, {Sanjay
  Ghemawat}, {Ian Goodfellow}, {Andrew Harp}, {Geoffrey Irving}, {Michael
  Isard}, Jia, {Rafal Jozefowicz}, {Lukasz Kaiser}, {Manjunath Kudlur}, {Josh
  Levenberg}, {Dan Man\'e}, {Rajat Monga}, {Sherry Moore}, {Derek Murray},
  {Chris Olah}, {Mike Schuster}, {Jonathon Shlens}, {Benoit Steiner}, {Ilya
  Sutskever}, {Kunal Talwar}, {Paul Tucker}, {Vincent Vanhoucke}, {Vijay
  Vasudevan}, {Fernanda Vi\'egas}, {Oriol Vinyals}, {Pete Warden}, {Martin
  Wattenberg}, {Martin Wicke}, {Yuan Yu}, and {Xiaoqiang
  Zheng}]{martinabadi2015_TensorFlow}
{Mart\'in Abadi},; {Ashish Agarwal},; {Paul Barham},; {Eugene Brevdo},;
  {Zhifeng Chen},; {Craig Citro},; {Greg S. Corrado},; {Andy Davis},; {Jeffrey
  Dean},; {Matthieu Devin},; {Sanjay Ghemawat},; {Ian Goodfellow},; {Andrew
  Harp},; {Geoffrey Irving},; {Michael Isard},; Jia,~Y.; {Rafal Jozefowicz},;
  {Lukasz Kaiser},; {Manjunath Kudlur},; {Josh Levenberg},; {Dan Man\'e},;
  {Rajat Monga},; {Sherry Moore},; {Derek Murray},; {Chris Olah},; {Mike
  Schuster},; {Jonathon Shlens},; {Benoit Steiner},; {Ilya Sutskever},; {Kunal
  Talwar},; {Paul Tucker},; {Vincent Vanhoucke},; {Vijay Vasudevan},; {Fernanda
  Vi\'egas},; {Oriol Vinyals},; {Pete Warden},; {Martin Wattenberg},; {Martin
  Wicke},; {Yuan Yu},; {Xiaoqiang Zheng}, \emph{{{TensorFlow}}: Large-{{Scale
  Machine Learning}} on {{Heterogeneous Systems}}}; 2015\relax
\mciteBstWouldAddEndPuncttrue
\mciteSetBstMidEndSepPunct{\mcitedefaultmidpunct}
{\mcitedefaultendpunct}{\mcitedefaultseppunct}\relax
\EndOfBibitem
\bibitem[Chollet \latin{et~al.}(2015)Chollet, \latin{et~al.}
  others]{chollet2015_Keras}
Chollet,~F., \latin{et~al.}  Keras. 2015\relax
\mciteBstWouldAddEndPuncttrue
\mciteSetBstMidEndSepPunct{\mcitedefaultmidpunct}
{\mcitedefaultendpunct}{\mcitedefaultseppunct}\relax
\EndOfBibitem
\bibitem[Kingma and Ba(2017)Kingma, and Ba]{kingma2017_Adam}
Kingma,~D.~P.; Ba,~J. Adam: A {{Method}} for {{Stochastic Optimization}}.
  \emph{ArXiv14126980 Cs} \textbf{2017}, \relax
\mciteBstWouldAddEndPunctfalse
\mciteSetBstMidEndSepPunct{\mcitedefaultmidpunct}
{}{\mcitedefaultseppunct}\relax
\EndOfBibitem
\bibitem[Li \latin{et~al.}(2018)Li, Jamieson, DeSalvo, Rostamizadeh, and
  Talwalkar]{li2018_Hyperband}
Li,~L.; Jamieson,~K.; DeSalvo,~G.; Rostamizadeh,~A.; Talwalkar,~A. Hyperband: A
  {{Novel Bandit}}-{{Based Approach}} to {{Hyperparameter Optimization}}.
  \emph{ArXiv160306560 Cs Stat} \textbf{2018}, \relax
\mciteBstWouldAddEndPunctfalse
\mciteSetBstMidEndSepPunct{\mcitedefaultmidpunct}
{}{\mcitedefaultseppunct}\relax
\EndOfBibitem
\bibitem[Kubo \latin{et~al.}(2012)Kubo, Toda, and
  Hashitsume]{kubo2012statistical}
Kubo,~R.; Toda,~M.; Hashitsume,~N. \emph{Statistical physics II: nonequilibrium
  statistical mechanics}; Springer Science \& Business Media, 2012;
  Vol.~31\relax
\mciteBstWouldAddEndPuncttrue
\mciteSetBstMidEndSepPunct{\mcitedefaultmidpunct}
{\mcitedefaultendpunct}{\mcitedefaultseppunct}\relax
\EndOfBibitem
\bibitem[Frenkel and Smit(2002)Frenkel, and Smit]{frenkel2002understanding}
Frenkel,~D.; Smit,~B. \emph{Understanding molecular simulation: From algorithms
  to applications}; Computational Sciences Series; Elsevier (formerly published
  by Academic Press), 2002; Vol.~1; pp 1--638\relax
\mciteBstWouldAddEndPuncttrue
\mciteSetBstMidEndSepPunct{\mcitedefaultmidpunct}
{\mcitedefaultendpunct}{\mcitedefaultseppunct}\relax
\EndOfBibitem
\bibitem[Grandbois \latin{et~al.}(1999)Grandbois, Beyer, Rief,
  {Clausen-Schaumann}, and Gaub]{grandbois1999_How}
Grandbois,~M.; Beyer,~M.; Rief,~M.; {Clausen-Schaumann},~H.; Gaub,~H.~E. How
  {{Strong Is}} a {{Covalent Bond}}? \emph{Science} \textbf{1999}, \emph{283},
  1727--1730\relax
\mciteBstWouldAddEndPuncttrue
\mciteSetBstMidEndSepPunct{\mcitedefaultmidpunct}
{\mcitedefaultendpunct}{\mcitedefaultseppunct}\relax
\EndOfBibitem
\bibitem[Karl and Marktanner(2001)Karl, and Marktanner]{karl2001_experiment}
Karl,~N.; Marktanner,~J. Electron and hole mobilities in high purity anthracene
  single crystals. \emph{Mol. Cryst. Liq. Cryst. A} \textbf{2001}, \emph{355},
  149--173\relax
\mciteBstWouldAddEndPuncttrue
\mciteSetBstMidEndSepPunct{\mcitedefaultmidpunct}
{\mcitedefaultendpunct}{\mcitedefaultseppunct}\relax
\EndOfBibitem
\bibitem[Jurchescu \latin{et~al.}(2004)Jurchescu, Baas, and
  Palstra]{jurchescu2004effect}
Jurchescu,~O.~D.; Baas,~J.; Palstra,~T.~T. Effect of impurities on the mobility
  of single crystal pentacene. \emph{Appl. Phys. Lett.} \textbf{2004},
  \emph{84}, 3061--3063\relax
\mciteBstWouldAddEndPuncttrue
\mciteSetBstMidEndSepPunct{\mcitedefaultmidpunct}
{\mcitedefaultendpunct}{\mcitedefaultseppunct}\relax
\EndOfBibitem
\end{mcitethebibliography}


\end{document}